\documentclass[letterpaper, 10 pt, conference]{ieeeconf}  

\IEEEoverridecommandlockouts                              

\usepackage{graphics} 
\usepackage{epsfig} 
\usepackage{times} 
\usepackage{amsmath} 
\usepackage{amssymb}  
\usepackage{dsfont}

\providecommand{\lemmaname}{\textbf{Lemma}}

\newtheorem{assumption}{\protect\assumname}
\providecommand{\assumname}{\textbf{Assumption}}

\usepackage{cite}
\usepackage{balance}
\usepackage{tikzscale}
\usepackage{pgfplots}
\usepackage{tikz}
\usepackage{xcolor}
\usepackage{graphicx}
\usepackage{url}

\DeclareMathOperator*{\esssup}{ess\,sup}

\title{\LARGE \bf
Risk-Sensitive Motion Planning using Entropic Value-at-Risk
}

\author{Anushri Dixit, Mohamadreza Ahmadi, and Joel W. Burdick$^{1}$
\thanks{$^{1}$The authors are with Control and Dynamical Systems at California Institute of Technology, 1200 E. Calif. Blvd., MC 104-44, Pasadena, CA 91125 {\tt\small \{adixit, mrahmadi\}@caltech.edu, jwb@robotics.caltech.edu}}%
}

\begin{document}

\maketitle
\thispagestyle{empty}
\pagestyle{empty}

\begin{abstract}
We consider the problem of risk-sensitive motion planning in the presence of randomly moving obstacles. To this end, we adopt a model predictive control (MPC) scheme and pose the obstacle avoidance constraint in the MPC problem as a distributionally robust constraint with a KL divergence ambiguity set. This constraint is the dual representation of the Entropic Value-at-Risk (EVaR). Building upon this viewpoint, we propose an algorithm to follow waypoints and discuss its feasibility and completion in finite time. We compare the policies obtained using EVaR with those obtained using another common coherent risk measure, Conditional Value-at-Risk (CVaR), via numerical experiments for a 2D system. We also implement the waypoint following algorithm on a 3D quadcopter simulation.
\end{abstract}

\section{Introduction}

Emerging applications in robot path planning in unknown and partially known unstructured environments, such as search and rescue missions caused by natural disasters~\cite{nagatani2013emergency,seraj2020coordinated}, inspection of planetary terrains~\cite{husain2013mapping}, and exploration of urban underground environments~\cite{kolvenbach2020towards},  motivate the need for risk-sensitive path planning. In particular,  path planning in subterranean environments~\cite{rouvcek2019darpa} incurs higher risks due to  lack of Global Positioning System (GPS) signals, the absence of illumination,  decentralization, and unpredictable environment topologies~\cite{papachristos2019autonomous,mansouri2020unified} (see Fig. 1).


Motion planning risk can be quantified in multiple ways, such as chance constraints~\cite{ono2015chance,wang2020non}, exponential utility functions~\cite{koenig1994risk}, and distributional robustness~\cite{xu2010distributionally}. However, applications in autonomy and robotics require more ``nuanced assessments of risk''~\cite{majumdar2020should}. Artzner \textit{et. al.}~\cite{artzner1999coherent} characterized a set of natural properties that are desirable for a risk measure.  These {\em coherent risk measures} are widely used and accepted in finance and operations research, among other fields. 

CVaR is an important example of a coherent risk measure that has received significant attention in decision making problems, such as Markov decision processes (MDPs)~\cite{chow2015risk,chow2014algorithms,prashanth2014policy,bauerle2011markov}. For stochastic discrete-time dynamical systems, a MPC technique for a class of coherent risk objectives that admit polytopic representation was proposed in~\cite{singh2018framework}. These authors also proposed a Lyapunov condition for risk-sensitive exponential stability in the presence of discretely quantized process noise, but did not include constraints in their formulation. Measurement noise and moving obstacles were considered in~\cite{hakobyan2019risk}, wherein the authors devised an MPC-based scheme for path planning with CVaR safety constraints when a reference
trajectory is generated by RRT$^*$~\cite{karaman2011sampling}, and extended to a Wasserstein distributionally robust formulation in~\cite{hakobyan2020wasserstein}. Risk-sensitive obstacle avoidance has also been tackled through CVaR control barrier functions in \cite{ahmadi2020cvar} with application to bipedal robot locomotion. Moreover, a method based on stochastic reachability analysis was proposed in~\cite{chapman2019risk} to estimate a CVaR-safe set of initial conditions via the solution to an MDP. 


\begin{figure}[t]
\centerline{\includegraphics[width=70mm,scale=0.5]{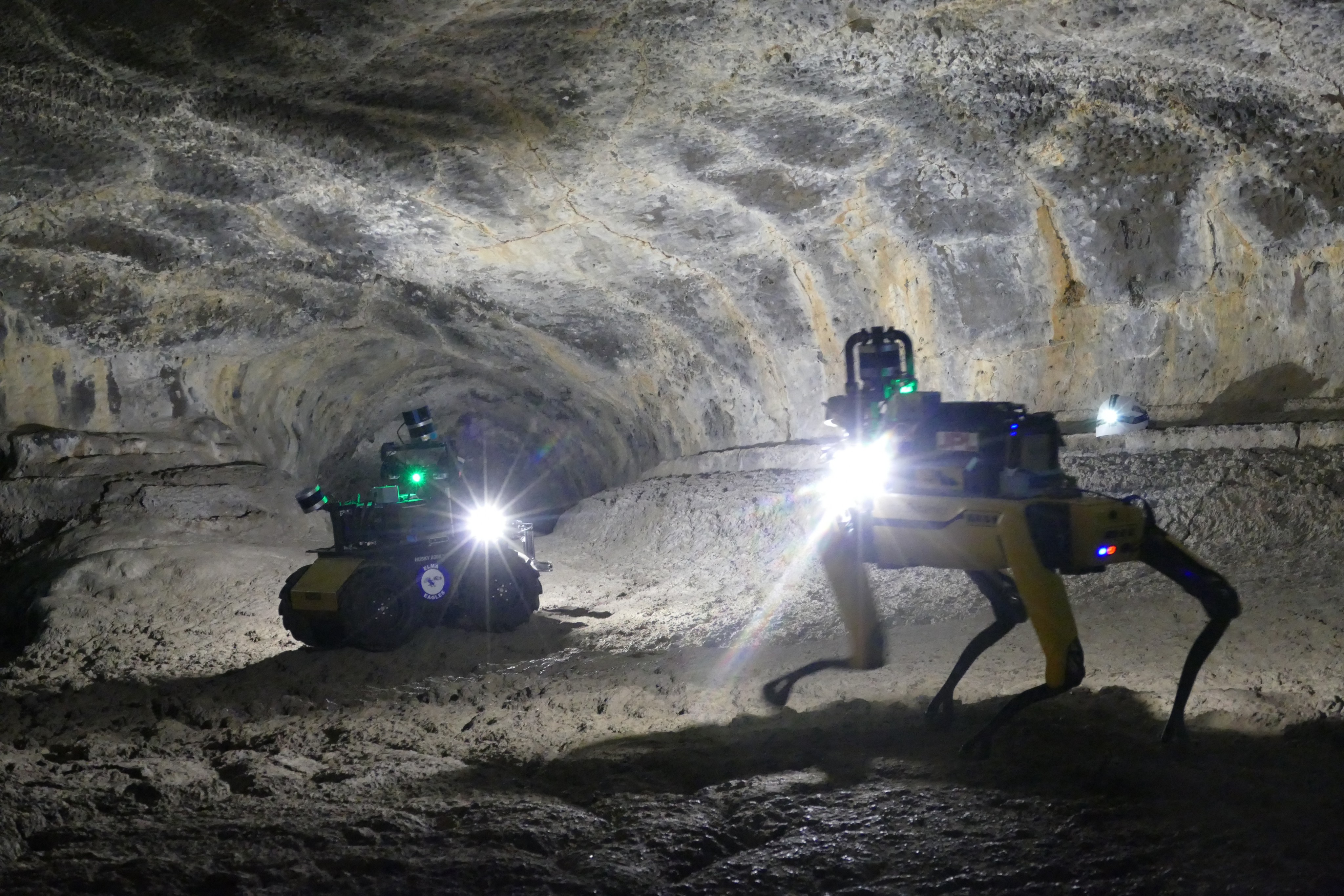}}
\caption{Spot and Husky robots
exploring a subterranean environment in Valentine Cave, Lava Beds National Monument, California. Avoidance of obstacles and other moving robots in unstructured environments incurs higher mission risk due to lack of global positioning~\cite{bouman2020autonomous}.}  \label{fig:waypoints}
\end{figure}


Despite the popularity of CVaR in risk-sensitive path planning, CVaR is hard to compute efficiently, even for the sum of arbitrary independent random variables~\cite{ahmadi2012entropic,ahmadi2012addendum}. In most cases, one has to approximate CVaR through sampling methods. Furthermore, CVaR only considers the average worst case performance beyond a threshold, while ignoring the performance before reaching that threshold~\cite{ahmadi2017analytical}. 

EVaR is the tightest upper bound on CVaR and Value-at-Risk (VaR) in the sense of the Chernoff inequality (and hence a more risk-sensitive measure)~\cite{ahmadi2012entropic,ahmadi2017analytical,ahmadi2019portfolio}. Nonetheless, it does not admit a polytopic representation and therefore methods such as the ones proposed in~\cite{singh2018framework} cannot be applied for risk-sensitive path planning. In \cite{Sopasakis2019}, risk-constrained and risk-averse optimal control, amenable to arbitrary coherent risk measures, is considered. The authors reformulate the optimal control optimization as a convex conic program. This formulation, however, does not consider the nonconvex, mixed-integer nature of the optimization problems that is often a result of obstacle avoidance constraints.
 
In this paper, we go beyond CVaR path planning and propose a framework for receding horizon path planning with risk-sensitive obstacle avoidance and guaranteed performance in terms of EVaR. We consider discrete-time systems and a class of randomly moving obstacles for which we reformulate the MPC optimization as a convex, mixed-integer program. This is done in three steps - first we write the EVaR constraint as a cone constraint, next we reformulate the obstacle avoidance constraints to obtain a mixed-integer relaxation, and lastly we add a discrete state that tracks whether the goal has been reached. This allows us to track waypoints in a way that guarantees feasibility and finite-time task completion. We elucidate the proposed method using two examples.

This paper is organized as follows. The next section briefly reviews some relevant facts on CVaR and EVaR. Section III presents the problem under study in this paper. In Section IV, we propose a reformulation based on convex mixed integer programming to solve the EVaR receding horizon path planning problem and discuss its feasibility properties. In Section V, we introduce an algorithm to follow waypoints using the aforementioned MPC optimization and prove its finite-time completion. Section VI illustrates the method via numerical experiments. Section VII concludes the paper.
\vspace{0.3cm}

\textbf{Notation: } We denote by $\mathbb{R}^n$ the $n$-dimensional Euclidean space, $\mathbb{R}_{\ge 0}$ the non-negative reals, and $\mathbb{N}_{\ge0}$ the set of non-negative integers. Throughout the paper, we use bold font to denote a vector and $(\cdot)^\top$ for its transpose, \textit{e.g.,} $\boldsymbol{a}=(a_1,\ldots,a_n)^\top$, with $n\in \{1,2,\ldots\}$. For vector $\boldsymbol{a}$, we use $\boldsymbol{a}\succeq (\preceq) \boldsymbol{0}$ to denote element-wise non-negativity (non-positivity) and $\boldsymbol{a}\equiv \boldsymbol{0}$ to show all elements of $\boldsymbol{a}$ are zero. For two vectors $a,b \in \mathbb{R}^n$, we denote their inner product by $\langle \boldsymbol{a}, \boldsymbol{b} \rangle$, \textit{i.e.,} $\langle \boldsymbol{a}, \boldsymbol{b} \rangle=\boldsymbol{a}^\top \boldsymbol{b}$. In the MPC problem, we refer to $\boldsymbol{x}(t+k|t)$ as $\boldsymbol{x}_k$. For a finite set $\mathcal{A}$, we denote its power set by $2^\mathcal{A}$.
For  a probability space $(\Omega, \mathcal{F}, \mathbb{P})$ and a constant $p \in [1,\infty)$, $\mathcal{L}_p(\Omega, \mathcal{F}, \mathbb{P})$ denotes the vector space of real valued random variables $X$ for which $\mathbb{E}|X|^p < \infty$. For two probability density functions $P(X)$ and $Q(X)$, $P \ll Q$ implies that $P$ is absolutely continuous with respect to $Q$, \textit{i.e.,} if $Q(X)=0$, then $P(X)=0$.


\section{Preliminaries}

This section reviews some results on CVaR and EVaR risk measures.

\begin{subsection}{Conditional Value-at-Risk}

For a given confidence level $\alpha \in (0,1)$, value-at-risk ($\mathrm{VaR}_{1-\alpha}$) denotes the $({1-\alpha})$-quantile value of the cost variable $X \in \mathcal{L}_p(\Omega, \mathcal{F}, \mathbb{P})$. $\mathrm{CVaR}_{1-\alpha}$ measures the expected loss in the $({1-\alpha})$-tail given that the particular threshold $\mathrm{VaR}_{1-\alpha}$ has been crossed. $\mathrm{CVaR}_{1-\alpha}$ is given by 
\begin{equation}
\begin{aligned}
    \mathrm{CVaR}_{1-\alpha}(X):=&\inf_{z \in \mathbb{R}}\mathbb{E}\Bigg[z + \frac{(X-z)^{+}}{1-\alpha}\Bigg], 
\end{aligned}
\end{equation}
 where $(\cdot)_{+}=\max\{\cdot, 0\}$. A value of $\alpha \simeq 0$ corresponds to a risk-neutral case; whereas, a value of $\alpha \to 1$ is rather a risk-averse case.  

\end{subsection}


\begin{subsection}{Entropic Value-at-Risk}

EVaR, derived using the Chernoff inequality for VaR, is the tightest upper bound for VaR and CVaR.  It was shown in~\cite{ahmadi2017analytical} that $\mathrm{EVaR}_{1-\alpha}$ and $\mathrm{CVaR}_{1-\alpha}$ are equal only if there are no losses ($X\to -\infty$) below the $\mathrm{VaR}_{1-\alpha}$ threshold. The $\mathrm{EVaR}_{1-\alpha}$ of random variable $X$ is given~by
\begin{equation} \label{eq:evar}
   \mathrm{EVaR}_{1-\alpha}(X):= \inf_{z > 0 }\Bigg[z^{-1}\ln\frac{\mathbb{E}[e^{Xz}]}{1-\alpha}\Bigg]
    = \sup_{Q\in\mathfrak{D}}\mathbb{E}_Q(X).
\end{equation}
%
Similar to $\mathrm{CVaR}_{1-\alpha}$, for $\mathrm{EVaR}_{1-\alpha}$, the limit $\alpha \to 0$ corresponds to a risk-neutral case; whereas, $\alpha \to 1$ corresponds to a risk-averse case. In fact, it was demonstrated in~\cite[Proposition 3.2]{ahmadi2012entropic} that $\lim_{{\alpha}\to 1} \mathrm{EVaR}_{{1-\alpha}}(X) = \esssup(X)$. 

A property of coherent risk measures is that they can be written as the worst-case expectation over a convex, bounded, and closed set of probability mass (or density) functions (pdf/pmf). This is the dual representation of a risk measure and the set is referred to as the risk envelope. For EVaR, the risk envelope $\mathfrak{D}$ for a continuous random variable with the pdf $P$ is defined as the epigraph of the KL divergence, given by,
\begin{multline}
    \mathfrak{D}:= \Big\{Q\ll P\, | \\ \, D_{KL}(Q||P):=\int\frac{dQ}{dP}\bigg(\ln\frac{dQ}{dP}\bigg)dP \leq -\ln(1-\alpha)\Big\}.
\end{multline}

\noindent $D_{KL}(Q||P)$ is the KL divergence between the two distributions, $Q$ and $P$. For some $x,y \in \mathbb{R}$, $D_{KL}(x||y)$ can be written in the form of the exponential cone, $K_{exp}$:
\begin{equation*}
    t\geq x\ln(x/y) \iff  (y, x, -t) \in K_{exp}.
\end{equation*}
Similarly, for a discrete random variable $X \in \{x_1, x_2,\dotsc, x_J\}$ with the pmf given by $p = [p(1), p(2), \dotsc, p(J)]^T$, where $p(j) = \mathbb{P}(X = x_j), j \in \{1,\dotsc, J\}$, the KL divergence is given as
\begin{equation*}
    D_{KL}(q||p) := \sum_{j=1}^{J}q(j)\ln\bigg(\frac{q(j)}{p(j)}\bigg), \quad q, p \in \Delta_J.
\end{equation*}
$\Delta_J$ is the probability simplex, $ \Delta_J:= \{q \in \mathbb{R}^J\,|\, q \geq 0, \,\sum_{j=1}^{J}q(j) = 1\}$. Hence, the epigraph of the KL divergence, and equivalently the EVaR risk envelope for a discrete distribution, is exponential cone representable \cite{kocuk2020conic} as
\begin{equation} \label{eq:KLcone}
\begin{aligned}
    \mathfrak{D}:= \Big\{& q \in \Delta_J \,|\,\exists \delta \in \mathbb{R}^J: \sum_{j=1}^{J}\delta(j) \leq -\ln\alpha, \\
    & \big(p(j), q(j), -\delta(j)\big) \in K_{exp},~ \, \forall j\in \{1, \dotsc, J\}\Big\}.
\end{aligned}
\end{equation}
\end{subsection}

\section{Problem Statement}
We consider a class of discrete-time systems given by
\begin{equation}\label{eq:sys}
\begin{aligned}
    \boldsymbol{x}(t+1) &= A\boldsymbol{x}(t) + B\boldsymbol{u}(t), \\
    \boldsymbol{y}(t) &= C\boldsymbol{x}(t) + D\boldsymbol{u}(t),
\end{aligned}
\end{equation}
where $\boldsymbol{x}(t) \in \mathbb{R}^{n_x}$, $\boldsymbol{y}(t) \in \mathbb{R}^{n_y}$, and $\boldsymbol{u}(t) \in \mathbb{R}^{n_u}$ are the system state, output, and controls at time $t$, respectively. 
We consider obstacles with index $l \in\mathcal{L}$ that can be approximated by a convex polytope defined by $m_l$ half-spaces in $\mathbb{R}^{n_x}$ 
\begin{equation} \label{eq: obs_def}
    \mathcal{O}_l = \{\boldsymbol{y} \in \mathbb{R}^{n_x} \,|\,\boldsymbol{c}_{i,l}^{T}\boldsymbol{y} \leq \boldsymbol{d}_{i,l}, \, i=1, \dots, m_l\}.
\end{equation}
We allow each polytopic obstacle $O_l$, $l \in \mathcal{L}$, centered at $\boldsymbol{a}_l$ at time $t$ to move randomly. That is, the point set defining obstacle $O_l$, $l \in \mathcal{L}$, at $t+k$ can be written as a random rotation $R_l(t+k)$ and random translation $w_l(t+k)$ of the $l$th obstacle  $\mathcal{O}_l$ from time $t$ to $t+k$ as described below
\begin{equation}\label{eq:obs}
\begin{aligned}
    \mathcal{O}_l(t+k) &= R_l(t+k)\mathcal{O}_l(t) + \boldsymbol{w}_l(t+k)\\
    &= \bigg\{\boldsymbol{y}(t+k) = R_l(t+k)(\boldsymbol{y}(t) - \boldsymbol{a}_l) + \boldsymbol{a}_l + \\ &\qquad \boldsymbol{w}_l(t+k) \, | \, \boldsymbol{c}_{i,l}^{T}\boldsymbol{y}(t) \leq \boldsymbol{d}_{i,l}, \, i=1, \dots, m_l\bigg\} \\
    &= \bigg\{\boldsymbol{y}(t+k) \, | \,\boldsymbol{c}_{i,l}^{T}\Big(R_l^{-1}(t+k)\big(\boldsymbol{y}(t+k)  - \boldsymbol{a}_l \\  & \qquad  -\boldsymbol{w}_l(t+k)\big) + \boldsymbol{a}_l\Big)  \leq \boldsymbol{d}_{i,l}, \, i=1, \dots, m_l\bigg \}.
\end{aligned}
\end{equation}
\begin{assumption}
\textit{The random rotations and translations are given by a joint probability distribution such that the sample space of this joint distribution has cardinality $J$, i.e., $\Omega_l = \{(R_l^1, \boldsymbol{w}_l^1), \dotsc, (R_l^J, \boldsymbol{w}_l^J)\}$. A random rotation and translation is picked from this set with pmf given by $p_l = [p(1), p(2), \dotsc, p(J)]^T$. For this distribution, we also define the index set $\mathcal{J} = \{1, \dotsc, J\}$.}
\end{assumption}
\vspace{0.3cm}

The safe set is defined as the region outside of the polytopic obstacles
\begin{equation}\label{eq:safeset_def}
\begin{aligned}
   \mathcal{S}_l(t) &= \mathbb{R}^{n_y} \backslash \mathcal{O}_l(t) \\
     &= \big\{\mathbf{y(t)}\,|\, \exists i \in \{1, \dots, m_l \},  c_{i,l}^T\mathbf{y(t)} \geq d_{i,l}\big\}.
     \end{aligned}
\end{equation}
\noindent For obstacle avoidance, we aim to minimize the distance to the safe set, which is given by
\begin{equation} \label{eq:zeta}
    \zeta(\mathcal{S}_l(t)) = \text{dist}(\boldsymbol{y}(t), \mathcal{S}_l(t)):= \min_{\boldsymbol{z} \in \mathcal{S}_l(t)} ||\boldsymbol{y}(t) - \boldsymbol{z}||.
\end{equation}

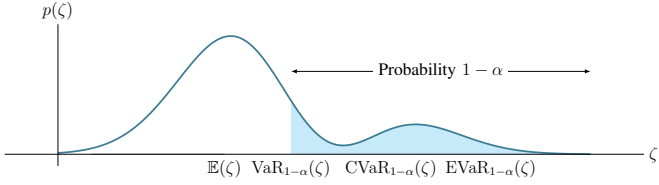
\begin{figure} 
\centering{
    \resizebox{0.5\textwidth}{!}{
\begin{tikzpicture}
\tikzstyle{every node}=[font=\large]
\pgfmathdeclarefunction{gauss}{2}{%
  \pgfmathparse{1000/(#2*sqrt(2*pi))*((x-.5-8)^2+.5)*exp(-((x-#1-6)^2)/(2*#2^2))}%
}

\pgfmathdeclarefunction{gauss2}{3}{%
\pgfmathparse{1000/(#2*sqrt(2*pi))*((#1-.5-8)^2+.5)*exp(-((#1-#1-6)^2)/(2*#2^2))}
}

\begin{axis}[
  no markers, domain=0:16, range=-2:8, samples=200,
  axis lines*=center, xlabel=$\zeta$, ylabel=$p(\zeta)$,
  every axis y label/.style={at=(current axis.above origin),anchor=south},
  every axis x label/.style={at=(current axis.right of origin),anchor=west},
  height=5cm, width=17cm,
  xtick={0}, ytick=\empty,
  enlargelimits=true, clip=false, axis on top,
  grid = major
  ]
  \addplot [fill=cyan!20, draw=none, domain=7:15] {gauss(1.5,2)} \closedcycle;
  \addplot [very thick,cyan!50!black] {gauss(1.5,2)};
 
 \pgfmathsetmacro\valueA{gauss2(5,1.5,2)}
 \draw [gray] (axis cs:5,0) -- (axis cs:5,\valueA);
  \pgfmathsetmacro\valueB{gauss2(10,1.5,2)}
  \draw [gray] (axis cs:4.5,0) -- (axis cs:4.5,\valueB);
    \draw [gray] (axis cs:10,0) -- (axis cs:10,\valueB);
 
 
 \draw [gray] (axis cs:1,0)--(axis cs:5,0);
\node[below] at (axis cs:7.0, -0.1)  {$\mathrm{VaR}_{1-\alpha}(\zeta)$}; 
\node[below] at (axis cs:5, -0.1)  {$\mathbb{E}(\zeta)$}; 
\node[below] at (axis cs:10, -0.1)  {$\mathrm{CVaR}_{1-\alpha}(\zeta)$}; \node[below] at (axis cs:13, -0.1)  {$\mathrm{EVaR}_{1-\alpha}(\zeta)$};
\draw [yshift=2cm, latex-latex](axis cs:7,0) -- node [fill=white] {Probability~$1-\alpha$} (axis cs:16,0);
\end{axis}
\end{tikzpicture}
}
\caption{Comparison of the mean, VaR, and CVaR for a given confidence $\alpha \in (0,1)$. The axes denote the values of the stochastic variable $\zeta$, i.e., the minimum distance to the safe set as defined in~\eqref{eq:zeta}, and with pdf $p(\zeta)$. The shaded area denotes the $\%(1-\alpha)$ of the area under $p(\zeta)$. If the goal is to minimize $\zeta$, using $\mathbb{E}(\zeta)$ as a performance measure is misleading because tail events with low probability of occurrence are ignored. VaR gives the value of $\zeta$ at the $(1-\alpha)$-tail of the distribution. But, it ignores the values of $\zeta$ with probability below $1-\alpha$.  CVaR is the average of the values of VaR with probability less than $1-\alpha$ (average of the worst-case values of $\zeta$ in the $1-\alpha$ tail of the distribution). Note that $\mathbb{E}(\zeta) \le \mathrm{VaR}_{1-\alpha}(\zeta) \le \mathrm{CVaR}_{1-\alpha}(\zeta) \le \mathrm{EVaR}_{1-\alpha}(\zeta)$. Hence,  $\mathrm{EVaR}_{1-\alpha}(\zeta)$ is a more risk-sensitive measure.}
}
\label{fig:varvscvar}
\end{figure}

Our goal is to minimize the risk of collision with the randomly moving obstacles by evaluating the EVaR of the distance from the probabilistic safe set (which is just the complement of the obstacle set) and constraining it to be below a certain threshold, $\epsilon_l$, i.e.,
\begin{equation}\label{eq:risksafety}
    \mathrm{EVaR}_{1-\alpha}\big[\zeta(\mathcal{S}_l(t))] \leq \epsilon_l, \quad  \forall l \in \mathcal{L}.
\end{equation} 

The obstacle avoidance constraint (\ref{eq:risksafety}) is an EVaR safety  constraint with confidence level $\alpha$ (see Fig. 2 for an illustrative comparison with VaR, CVaR and statistical mean) and risk tolerance (also referred to as risk threshold in this paper) $\epsilon_l$ for each obstacle~$l \in \mathcal{L}$. {\color{black} Note that this implies that we allow the EVaR of the distance from the safe set to be at most $\epsilon_l$ with $1-\alpha$ worst realizations.}
We are now ready to present the problem we are interested in solving in this paper.
\vspace{0.3cm}

\begin{problem}
\textit{
Consider the discrete-time system given by (\ref{eq:sys}) and the randomly moving obstacles $O_l$, $l \in \mathcal{L}$, as defined in~\eqref{eq: obs_def} and~\eqref{eq:obs}. Given an initial condition $x_0 \in \mathbb{R}^{n_x}$, a goal set $\mathcal{X}_f \subset \mathbb{R}^{n_x}$, state constraints $\mathcal{X}\subset \mathbb{R}^{n_x}$, control constraints $\mathcal{U}\subset \mathbb{R}^{n_u}$, an immediate convex cost function $r:\mathbb{R}^{n_x} \times \mathbb{R}^{n_u}\to \mathbb{R}_{\ge0}$, a horizon $K \in \mathbb{N}_{\ge 0}$, and risk tolerances $\epsilon_l$, $l \in \mathcal{L}$, for each obstacle, compute the receding horizon controller $\{u_k \}_{k=0}^{K-1}$ such that $x(K) \in \mathcal{X}_f$ while satisfying the risk-sensitive safety constraints~\eqref{eq:risksafety}, i.e., the solution to the following optimization~\begin{subequations}\label{eq:mpc1}
\begin{align}
\begin{split}
\min_{u} \quad &J(x(t), \boldsymbol{u}) := \sum_{k=0}^{K-1}r(\boldsymbol{x}_k, \boldsymbol{u}_k) \quad
\end{split}\\
\begin{split}\label{eq:dyn1}
 \textrm{s.t.} \quad &\boldsymbol{x}_{k+1} = A\boldsymbol{x}_k + B\boldsymbol{u}_k
\end{split}\\
\begin{split}\label{eq:dyn2}
&\boldsymbol{y}_k = C\boldsymbol{x}_k + D\boldsymbol{u}_k 
\end{split}\\
\begin{split}\label{eq:stcon}
&\boldsymbol{x}_k \in \mathcal{X}, \, \boldsymbol{u}_k \in \mathcal{U}, 
\end{split}\\
\begin{split}\label{eq:evarcon}
&\mathrm{EVaR}_{1-\alpha}\big[\zeta(\mathcal{S}_l(t+k))\big] \leq \epsilon_l, \forall l \in \mathcal{L}, 
\end{split}\\
\begin{split}
x_K  \in \mathcal{X}_F.
\end{split}\\
\begin{split}\label{eq:ic}
&\boldsymbol{x}_0 = \boldsymbol{x}(t), 
\end{split}
\end{align}
\end{subequations}
}
\end{problem}
\vspace{0.3cm}

Note that although the obstacles $\mathcal{O}_l$ are assumed to be represented by convex polytopes \eqref{eq: obs_def}, the safe set $\mathcal{S}_l(t+k)$ given in~\eqref{eq:safeset_def} is nonconvex. Hence, the minimum distance to $\mathcal{S}_l(t+k)$, $\zeta(\mathcal{S}_l(t+k))$, is also nonconvex. Therefore, the risk-sensitive safety constraint~\eqref{eq:evarcon} is a nonconvex constraint in the decision variable $u$, which renders optimization problem~\eqref{eq:mpc1} nonconvex as well. 

The next section will reformulate \eqref{eq:evarcon} as a cone constraint in order to obtain a convex mixed-integer relaxation of \eqref{eq:mpc1}, which yields locally optimal solutions to \eqref{eq:mpc1}. Nonetheless, every such locally optimal solutions satisfies the constraints of optimization~\eqref{eq:mpc1} including the risk-sensitive safety constraint~\eqref{eq:evarcon}. 




\section{EVaR Receding Horizon Planning}

This section breaks down the MPC optimization into three parts. First, we rewrite the EVaR constraint in the more tractable form of a cone constraint. Second, we reformulate the nonconvex safe set as a set of disjunctive inequalities that can be relaxed using binary variables. Lastly, we add a discrete state $\psi$ that signals task completion and allows us to prove feasibility of the MPC optimization. The resulting optimization is a convex mixed-integer program. 

\subsection{EVaR Constraint Reformulation}

We reformulate the EVaR safety constraint to a cone constraint. 


\begin{lemma} \label{lemma}
\textit{Let Assumption 1 hold, then the L.H.S. of constraint \eqref{eq:evarcon} is equivalent to
\begin{equation} \label{eq:evar_reformulation}
\begin{aligned}
    \min_{s_l, v_l, z_l, h_{l,k}} \, &\eta_l - \beta_l\ln\alpha + \sum_{j=1}^{J}p_l(j)s_l(j) \qquad \\
    \textrm{s.t.} \quad &  \eta_l \in \mathbb{R}, \, \beta_l \in \mathbb{R}_{\ge0}, \\
    & \eta_l-v_l(j) \geq h_{l,k},& \, \forall j \in \mathcal{J},\\
    &  \beta_l + z_l(j) = 0, & \, \forall j \in \mathcal{J},\\
    &  \big(s_l(j), v_l(j), z_l(j)\big) \in (K_{exp})_*, &  \forall j \in \mathcal{J},\\
    & \boldsymbol{y}_k +\frac{\boldsymbol{c}_{i,l}}{||\boldsymbol{c}_{i,l}||}h_{l,k} \in \mathcal{S}_l^j(t+k), & \forall l \in \mathcal{L}, j \in \mathcal{J},
\end{aligned}
\end{equation}
where $(K_{exp})_*$ is the dual of the exponential cone.}
\end{lemma}

\vspace{0.2cm}
\begin{proof}
We begin by finding the distance of $\boldsymbol{y}_k$ from the safe set, given by $\zeta(\mathcal{S}_l(t+k))$. To this end, we define a set of variables $h_{l,k}\ge 0$, $l \in \mathcal{L}$ and $k=0,\ldots,K-1$ satisfying

\begin{equation}\label{eq:safeset1}
    \boldsymbol{y}_k +\frac{\boldsymbol{c}_{i,l}}{||\boldsymbol{c}_{i,l}||}h_{l,k} = \boldsymbol{z}
\end{equation}
 $\forall j \in \mathcal{J}, \forall k \in \{0, \dotsc, K-1\}$ with $i \in \{1, \dotsc, m_l\}$, which is the distance from every $\boldsymbol{y}_k$ to a point $\boldsymbol{z} \in \mathcal{X}$. If $\boldsymbol{z} \in \mathcal{S}_l^j(t+k)$, then  minimizing $h_{l,k} $ gives us the line segment connecting $ \boldsymbol{y}_k$ and $\boldsymbol{z}$, which is the minimum distance to the set $ \mathcal{S}_l^j(t+k)$. Therefore, we obtain
 \begin{multline}\label{eq:safeset1}
    \zeta(\mathcal{S}_l(t+k)) = \min_{\boldsymbol{z} \in \mathcal{S}_l(t+k)} ||\boldsymbol{y}(t+k) - \boldsymbol{z}|| \\ = \left \lbrace\begin{matrix} \min_{h_{l,k}} & h_{l,k} \qquad \\
     \textrm{s.t.}&\boldsymbol{y}_k +\frac{\boldsymbol{c}_{i,l}}{||\boldsymbol{c}_{i,l}||}h_{l,k} \in \mathcal{S}_l^j(t+k), \,~~ \forall j \in \mathcal{J}, \end{matrix} \right.
 \end{multline}
 and define $h_{l,k}^*$ as the solution to~\eqref{eq:safeset1}.
 
Next, substitute the dual form of EVaR from \eqref{eq:evar} into the L.H.S. of \eqref{eq:evarcon}. Then replace the risk envelope $\mathfrak{D}$ with the exponential cone representation to yield a discrete probability distribution given by \eqref{eq:KLcone}. That is, $\mathrm{EVaR}_{1-\alpha}(\zeta(\mathcal{S}_l(t+k))) = \max_{Q\in\mathfrak{D}}\mathbb{E}_Q\big[\zeta(\mathcal{S}_l(t+k))\big]=\max_{Q\in\mathfrak{D}}\mathbb{E}_Q\big[h_{l,k}^*\big]$, where in the last equality we used~\eqref{eq:safeset1}.   Thus, we have the following exponential cone program for computing EVaR
\begin{equation}\label{eq:dual}
\begin{aligned}
    \max\limits_{q, \delta} \quad &\sum\limits_{j=1}^{J} h_{l,k}^* q(j) \\
    \textrm{s.t.} \quad & \sum_{j=1}^J q(j) = 1,\\
    & \sum_{j=1}^{J} \delta(j) \leq -\ln\alpha,\\
    & \big((p_l(j), q(j), -\delta(j)\big) \in K_{exp},~~ \forall j \in \mathcal{J},  \\
    & q(j) \in \mathbb{R}_{\ge 0}, \, \delta(j) \in \mathbb{R}, \phantom{--iiii}\forall j \in \mathcal{J}.
    \end{aligned}
\end{equation}

%
  The dual of the above maximization problem is given by~\cite{kocuk2020conic, Sopasakis2019}:
\begin{equation} \label{eq:dualofdual}
\begin{aligned}
    \min_{s_l, v_l, z_l} \, &\eta_l - \beta_l\ln\alpha + \sum_{j=1}^{J}p_l(j)s_l(j) \qquad \\
    \textrm{s.t.} \quad & \eta_l-v_l(j) \geq h_{l,k}^*,& \, \forall j \in \mathcal{J},\\
    &  \beta_l + z_l(j) = 0, & \, \forall j \in \mathcal{J},\\
    &  \eta_l \in \mathbb{R}, \, \beta_l \in \mathbb{R}_{\ge0}, \\
    &  \big(s_l(j), v_l(j), z_l(j)\big) \in (K_{exp})_*, &  \forall j \in \mathcal{J},
\end{aligned}
\end{equation}


\noindent where $\boldsymbol{s, v, z}$ are the dual variables.

We conclude that \eqref{eq:dualofdual} and \eqref{eq:dual} are equivalent because strong duality holds by Slater's condition \cite{2004boyd}. Slater's condition is satisfied by showing \textit{strict feasibility} for \eqref{eq:dual}, i.e., there exists a feasible solution to \eqref{eq:dual} such that the inequality constraints hold with strict inequalities. One such solution is when $q(j) = p(j), \,\delta(j) < 0, \, \forall j\in \mathcal{J}$.

Finally, substituting minimization problem~\eqref{eq:safeset1} for $h^*_{l,k}$ in  optimization \eqref{eq:dualofdual} gives  \eqref{eq:evar_reformulation}. \end{proof}
\vspace{0.3cm}
with additional variables in the optimization: $\boldsymbol{s_l, v_l, z_l},h_{l,k}$

Utilizing the fact that $\mathrm{EVaR}_{1-\alpha}(\mathcal{S}_l(t+k))$ can be written as a minimization over the variables $\boldsymbol{s_l, v_l, z_l}, h_{l,k}$, we can return to our original MPC problem~\eqref{eq:mpc1} and simplify it as a one-layer optimization.
\vspace{0.3cm}
\begin{theorem}
\textit{Consider the MPC optimization given by~(\ref{eq:mpc1}) with confidence level $\alpha$ and risk tolerances $\epsilon_l$, $l \in \mathcal{L}$. If Assumption 1 holds, then \eqref{eq:mpc1} is equivalent to a minimization over $\mathcal{V} = \{\boldsymbol{u,s_l,v_l,z_l},h_{l,k}\}$ given by}
\begin{subequations}\label{eq:mpc2}
\begin{align}
\begin{split}
\min_{\mathcal{V}} \quad &J(\boldsymbol{x}(t), \boldsymbol{u}) := \sum_{k=0}^{K-1}r(\boldsymbol{x}_k, \boldsymbol{u}_k) 
\end{split}\\
\begin{split}
\textrm{s.t.} \quad & \eta_l - \beta_l\ln\alpha + \sum_{j=1}^{J}p(j)s(j) \leq \epsilon_l \, \forall l \in \mathcal{L},
\end{split}\\
\begin{split}
& \eta_l-v_l(j) \geq  h_{l, k} \, \phantom{-----ii}\forall j \in \mathcal{J},l \in \mathcal{L},
\end{split}\\
\begin{split} \label{eq:safeset}
& \boldsymbol{y}_k +\frac{\boldsymbol{c}_{i,l}}{||\boldsymbol{c}_{i,l}||}h_{l,k} \in \mathcal{S}_l^j(t+k) \phantom{-} \forall l \in \mathcal{L},
\end{split}\\
\begin{split}
& \beta_l + z_l(j) = 0  \,\phantom{------ii}  \forall j\in \mathcal{J},l \in \mathcal{L},
\end{split}\\
\begin{split}
& \eta_l \in \mathbb{R}, \, \beta_l, h_{l,k} \in \mathbb{R}_{\geq 0} \,\phantom{---ii} \forall l \in \mathcal{L},
\end{split}\\
\begin{split}
& \big(s_l(j), v_l(j), z_l(j)\big) \in (K_{exp})_* \,\,\forall j\in \mathcal{J},l \in \mathcal{L},
\end{split}\\
\begin{split}
(\ref{eq:dyn1}),(\ref{eq:dyn2}),(\ref{eq:ic}), (\ref{eq:stcon}).
\end{split}
\end{align}
\end{subequations}
\end{theorem}
\vspace{0.3cm}
\begin{proof}
We can substitute the result from Lemma~\ref{lemma} in~\eqref{eq:mpc1} to get
\begin{subequations}\label{eq:mpc_minmin}
\begin{align}
\begin{split}
\min_{\boldsymbol{u}} \quad &J(\boldsymbol{x}(t), \boldsymbol{u}) := \sum_{k=0}^{K-1}r(\boldsymbol{x}_k, \boldsymbol{u}_k) \quad \end{split}\\
\begin{split}
\textrm{s.t.} \quad & (\ref{eq:dyn1}),(\ref{eq:dyn2}),(\ref{eq:ic}), (\ref{eq:stcon}),
\end{split}\\
\begin{split}
\eqref{eq:evar_reformulation} \leq \epsilon_l, \quad l \in \mathcal{L}.
\end{split}
\end{align}
\end{subequations}

Suppose we have an optimal solution to \eqref{eq:mpc_minmin} given by $\boldsymbol{u}^*$. As \eqref{eq:mpc_minmin} is feasible, its constraints must be satisfied; this implies the inner minimization \eqref{eq:evar_reformulation} must also be feasible (with solution $(\boldsymbol{s_l^*, v_l^*, z_l^*}, h_{l,k}^*)$). Hence, $(\boldsymbol{u^*,s_l^*, v_l^*, z_l^*}, h_{l,k}^*)$ must also be a feasible solution to \eqref{eq:mpc2} and give the same objective value. Conversely, consider the optimal solution to \eqref{eq:mpc2} to be given by $(\boldsymbol{u^*,s_l^*, v_l^*, z_l^*}, h_{l,k}^*)$. This $\boldsymbol{u}^*$ must be feasible for \eqref{eq:mpc_minmin} and gives the same objective value. Hence, the above optimization \eqref{eq:mpc_minmin} is equivalent to the one-layer optimization~\eqref{eq:mpc2}. 
\end{proof}

\subsection{Mixed-Integer Reformulation of the MPC optimization}

This subsection frames the nonconvex safe set as a set of disjunctive inequalities. These inequalities are incorporated in our optimization by introducing a set of binary variables and using the Big-M relaxation~\cite{vecchietti_modeling_2003}.

The safe set~\eqref{eq:safeset_def} is defined as the region outside the obstacle $l$. Given that an obstacle has rotated and translated by $R_l(t+k)$ and $\boldsymbol{w}_l(t+k)$ between times $t$ and $t+k$, we can write the safe set at $t+k$ as the region outside $\mathcal{O}_l(t+k)$ described in \eqref{eq:obs}. It can equivalently be expressed as a result of the rotation and translation of the safe set itself from $t$ to $t+k$ 
\begin{equation}
\begin{aligned}
    \mathcal{S}_l^j(t+k) & = \mathbb{R}^{n_y} \backslash \mathcal{O}_l^j(t+k) \\
    & = R_l^j(t+k)\mathcal{S}_l(t) + \boldsymbol{w}_l^j(t+k).
\end{aligned}
\end{equation}

In (\ref{eq:safeset}), $\mathcal{S}_l^j(t+k)$ is a nonconvex set. For some obstacle~$l \in \mathcal{L}$, we can be rewrite \eqref{eq:safeset} as
\begin{equation*}
    R_l^j(t+k)^{-1}\bigg(\boldsymbol{y}_k +\frac{\boldsymbol{c}_{i,l}}{||\boldsymbol{c}_{i,l}||}h_{l,k} - \boldsymbol{w}_l^j(t+k)\bigg) \in \mathcal{S}_l^j(t).
\end{equation*}

Given that the obstacles are convex polygons of the form~\eqref{eq: obs_def}, we write the safe region as the union of regions outside of the halfspaces that define an obstacle as follows
{\small
\begin{multline}\label{eq:disjunction}
   \bigvee_{i=1}^{m_l} \boldsymbol{c}_{i,l}^T \Bigg[ R_l^j(t+k)^{-1}\bigg(\boldsymbol{y}_k +\frac{\boldsymbol{c}_{i,l}}{||\boldsymbol{c}_{i,l}||}h_{l,k} \\- \boldsymbol{w}_l^j(t+k) - \boldsymbol{a}_l\bigg) + \boldsymbol{a}_l \Bigg] \geq d_{i,l}.
\end{multline}}
The above disjunctive inequalities, however, are hard to enforce. To overcome this difficulty, we relax the constraint using a Big-M reformulation. The reformulation converts the disjunctive inequalities into a set of constraints described using binary variables, $\gamma_i \in \{0,1\}$ and a large positive constant $M$. The value of $M$ depends on the bounds on $h_{l,k}$ (determined from the size of obstacle $l$) and $\boldsymbol{y}_k$ (dependent on the state and control inputs). It can be computed using linear programming. The Big-M relaxation of \eqref{eq:disjunction} is as follows

 \begin{subequations} \label{eq:bigM}
   \begin{align}
 \begin{split}\label{eq:bigM1} 
     \boldsymbol{c}_{i,l}^T \Bigg[ R_l^j(t+k)^{-1}&\bigg(\boldsymbol{y}_k +\frac{\boldsymbol{c}_{i,l}}{||\boldsymbol{c}_{i,l}||}h_{l,k} - \boldsymbol{w}_l^j(t+k) - \boldsymbol{a}_l\bigg)  + \boldsymbol{a}_l \Bigg] \\& \geq d_{i,l} - M\gamma_i, \qquad  \, \forall i \in \{1, \dotsc m_l \},
     \end{split}\\
    \begin{split}
        \sum_{i = 1}^{m_l} \gamma_i \leq m_l - 1.
    \end{split}
   \end{align}
 \end{subequations}
 

\subsection {Task Completion}
 In order to steer the system to the target region in finite time, we follow the footsteps of \cite{richards2003robustFeasibility} and define a new discrete state $\psi \in \{ 0, 1\}$, such that $\psi = 0$ implies that the task has been completed at an earlier step or at the current step and $\psi = 1$ means that the task has not yet been completed. The update equation of $\psi$ is then given by
\begin{equation}\label{eq:taskCompletion}
    \psi_{k+1} = \psi_{k} - \mu_{k},
\end{equation}
\noindent where ${\mu}_{k} \in \{ 0, 1\}$ is a discrete input. 
The goal to drive the system to $(\boldsymbol{x}_{des}, \boldsymbol{u}_{des})$ (this desired position can be replaced by a region as well), is incorporated in the following additional constraints

\begin{equation}\label{eq:terminalState}
\begin{aligned}
    \boldsymbol{x}_{k+1} - \boldsymbol{x}_{des} &\leq \mathds{1}{M}(1 - \mu_{k}), \\
    -(\boldsymbol{x}_{k+1} - \boldsymbol{x}_{des}) &\leq -\mathds{1}{M}(1 - \mu_{k}).
\end{aligned}
\end{equation}

Here ${\mu}_{k} = 1$ if the task of reaching the goal is completed at time step $t+k+1$. Equation \eqref{eq:taskCompletion} implies that $\psi$ jumps from $1 \rightarrow 0$, signaling completion of the task. After the task completion, all other MPC problem constraints can be relaxed by adding the term $M(1 - \psi_{k})$ to them, i.e., any constraints of the form $C_1\boldsymbol{\nu}_k + C_2\gamma_i + C_3 \geq 0$ are modified to $C_1\boldsymbol{\nu}_k + C_2\gamma_i  + C_3 + \mathds{1}M(1 - \psi_{k})\geq 0, \, \forall i, k$ where $\boldsymbol{\nu}_k = [\boldsymbol{u_k,x_k,y_k,s_k,v_k,z_k,h_k} ,\eta_k,\beta_k] $. We also add the following terminal constraint at time $t + K$ to ensure that the task is completed
\begin{equation} \label{eq:terminalConst}
    \psi_{K} = 0.
\end{equation}

Note that the discrete state $\psi$ need not be a binary variable as long as we enforce the constraint,
\begin{equation}\label{eq:discreteStateConst}
    0 \leq \psi_{k} \leq 1, \quad k=1,2,\ldots,K.
\end{equation}

The MPC objective function is then modified as
\begin{equation}
    \min_{\mathcal{V}} \quad J(t) := \sum_{k=0}^{K-1}\big(r(\boldsymbol{u}_k) + \psi_k\big),
\end{equation}
where $\mathcal{V} = \{\boldsymbol{u,v,h,s,z,\mu,\gamma}\}$ and $r(\boldsymbol{u}_k)$ is a convex function of the control input with $r(0) = 0$.

The MPC optimization \eqref{eq:mpc1} has the following convex mixed integer relaxation,
 \begin{subequations}\label{eq:mpc3} {\small
\begin{align}
\begin{split}
    \min_{\mathcal{V}} \quad & J(t) := \sum_{k=0}^{K-1}\big(r(\boldsymbol{u}_k) + \psi_k\big) 
\end{split}\\
\textrm{s.t.} \quad \begin{split}
& \eta_{l,k} - \beta_{l,k}\ln\alpha + \sum_{j=1}^{J}p_{l}(j)s_{l,k}(j) \leq \epsilon_l  + M_k,
\end{split}\\
\begin{split}
& \eta_{l,k}-v_{l,k}(j)+ M_k \geq  h_{l, k},
\end{split}\\
\begin{split}
& \text{L.H.S.}\eqref{eq:bigM1} + M_k\geq d_{i,l} - M\gamma_{i,l,k}(j),
\end{split}\\
\begin{split}
    \sum_{i = 1}^{m_l} \gamma_{i,l,k}(j) \leq m_l - 1 + M_k, 
\end{split}\\
\begin{split}
& -M_k \leq \beta_{l,k} + z_{l,k}(j) \leq M_k,
\end{split}\\
\begin{split}
& \beta_{l,k} +M_k,  \,h_{l,k}+ M_k \in \mathbb{R}_{\ge 0}, 
\end{split}\\
\begin{split}
& \big(s_{l,k}(j) + M_k, v_{l,k}(j), z_{l,k}(j)\big) \in (K_{exp})_*,  
\end{split}\\
\begin{split}
    (\boldsymbol{x}_k, \boldsymbol{u}_k, \psi_k) \in (\mathcal{X}, \mathcal{U}, 1) \cup (\mathbb{R}^n, \mathbb{R}^n, 0),
\end{split}\\
\begin{split}
& (\ref{eq:taskCompletion}), (\ref{eq:terminalState}), (\ref{eq:terminalConst}), (\ref{eq:discreteStateConst}), (\ref{eq:dyn1}),(\ref{eq:dyn2}),(\ref{eq:ic}),
\end{split}
\end{align}}
\end{subequations}
where $M_k = M(1 - \psi_{k})$ and $\mathcal{X}, \mathcal{U}$ are assumed to be convex sets. The constraints must hold $\forall k\in \{1, \dotsc, K-1\},\, l\in \mathcal{L},\, j \in \mathcal{J},$ and $ i \in \{1, \dotsc, m_l\}$. 

The above convex, mixed-integer relaxation of a nonconvex optimization problem will give us locally optimal solutions. 

\vspace{0.2cm}
\begin{proposition} \label{feasibility}
 \textit{   If the optimization \eqref{eq:mpc3} is feasible at time $t = 0$, it is feasible for future time steps.} 
\end{proposition}
\vspace{0.2cm}
\begin{proof}
Assume that the feasible solution to \eqref{eq:mpc3} at time $t$ is given by the input sequence $\{\boldsymbol{u}_0^*, \boldsymbol{u}_1^*, \dotsc , \boldsymbol{u}_{K-1}^*\}$ and the state sequence $\{(\boldsymbol{x}_0^*, \psi_0^*), (\boldsymbol{x}_1^*, \psi_1^*), \dotsc , (\boldsymbol{x}_K^*, \psi_K^*)\}$. Recall from the notation section that $\boldsymbol{x}_k := \boldsymbol{x}(t+k|t)$. Applying the first control input leads the system to the next state in the sequence $(\boldsymbol{x}_1^*, \psi_1^*)$, provided that the model of the actual system matches the one in \eqref{eq:mpc3}. The optimization is feasible at time $t+1$ if there exists a feasible input at time $t+K$ that does not violate constraints. Since $\psi_K^* = 0$ by virtue of the terminal constraint, all the constraints in the optimization are relaxed thereafter. Note that the state $\psi_K = 0$ is invariant due to \eqref{eq:taskCompletion} and \eqref{eq:discreteStateConst} and that $\mu_k = 0$ at all time after the task has been completed. Therefore, once the optimization constraints are relaxed, they will remain this way.
 
A control input $u_K = 0$ ensures that the optimization is feasible. At time $t+1$, a feasible solution to \eqref{eq:mpc3} is given by the control sequence $\{\boldsymbol{u}_1^*, \boldsymbol{u}_2^*, \dotsc , \boldsymbol{u}_{K-1}^*, 0\}$ and the state sequence $\{(\boldsymbol{x}_1^*, \psi_1^*), \dotsc ,(\boldsymbol{x}_K^*, 0), (A\boldsymbol{x}_{K}^*, 0)\}$. Hence, if the optimization \eqref{eq:mpc3} is feasible at time $t$, then it is feasible at time $t+1$. By extension, if the optimization \eqref{eq:mpc3} is feasible at time $t = 0$, it is feasible for all future time steps. 
\end{proof}

\section{Waypoint Following Algorithm}
MPC is often used as a tool to plan trajectories locally and it is given a reference trajectory or a set of waypoints from a higher-level global planner like A* or RRT ~\cite{Lopez2017, hakobyan2019cvar}. Let  $\{\boldsymbol{w}_1, \boldsymbol{w}_2, \dotsc, \boldsymbol{w}_N\}$ be a given a sequence of waypoints.

We call a waypoint $\boldsymbol{w}_{j+1}$ \textit{K-step reachable} from $\boldsymbol{w}_j$, if there exists a feasible solution to \eqref{eq:mpc3} with $\boldsymbol{x}_0 = \boldsymbol{w}_j$ and $\boldsymbol{x}_{K} = \boldsymbol{w}_{j+1}$.

\begin{algorithm}[t!]
\caption{Follow waypoints}\label{waypoint_alg}
\begin{algorithmic}
    \STATE Number of waypoints visited, $W = 0$
    \WHILE{$W < N$}
        \STATE Initialize $(\boldsymbol{x}_0, \psi_0) = (\boldsymbol{w}_{W}, 1)$
        \STATE Set desired goal $\boldsymbol{x}_{des} = \boldsymbol{w}_{W+1}$  
        \WHILE{$\psi_0 \neq 0$}
            \STATE Solve \eqref{eq:mpc3} to obtain policy $\{\boldsymbol{u}_0^*, \boldsymbol{u}_1^*, \dotsc , \boldsymbol{u}_{K-1}^*\}$ 
            \STATE Update $\boldsymbol{x}_0 = A\boldsymbol{x}_0 + B\boldsymbol{u}_0^*$ 
            \STATE Update $\psi_0 = \psi_0 - \mu_0$ 
            \IF{$\boldsymbol{x}_0 =\boldsymbol{x}_{des}$}
                \STATE $W = W + 1$
            \ENDIF
        \ENDWHILE
    \ENDWHILE
    \end{algorithmic}
\end{algorithm}


\vspace{0.2cm}
\begin{proposition}
 \textit{Assuming that the waypoint $\boldsymbol{w}_{j+1}$ is K-step reachable from $\boldsymbol{w}_j, \, \forall j \in \{1, \dotsc, N-1\}$, Algorithm \ref{waypoint_alg} gives a sequence of control inputs to move from $\boldsymbol{w}_0$ to $\boldsymbol{w}_N$ in finite time. }
\end{proposition}
\vspace{0.2cm}
\begin{proof}
Consider the simple case of starting from $\boldsymbol{w}_{0}$ and reaching $\boldsymbol{w}_{1}$, i.e., when we have exactly two waypoints. We implement Algorithm \ref{waypoint_alg} till $\psi_0 = 0$. Let $J_t^*$ be the cost of the MPC optimization \eqref{eq:mpc3} at time $t$. The feasible solution to \eqref{eq:mpc3} at $t$ is given by the input sequence $\{\boldsymbol{u}_{0}^*, \boldsymbol{u}_{1}^*, \dotsc , \boldsymbol{u}_{K-1}^*\}$ and the state sequence $\{(\boldsymbol{x}_{0}^*, \psi_{0}^*), (\boldsymbol{x}_{1}^*, \psi_{1}^*), \dotsc , (\boldsymbol{x}_{K}^*, \psi_{K}^*)\}$. At time $t+1$, the cost of the MPC optimization is $J_{t+1}^* \leq J_{t}^* - r(\boldsymbol{u}_{0}^*) - \psi_{0}^*$. This is true because we know from Proposition \ref{feasibility} that at time $t+ 1$, $\{\boldsymbol{u}_{1}^*, \dotsc , \boldsymbol{u}_{K-1}^*, 0\}$ is a feasible control input with $\psi(t+K|t+1) = 0$, i.e., $J_{t}^*$ will incur no additional cost from $u(t+K|t+1) = 0$ and $\psi(t+K+1|t+1) = 0$. Since $J_{t+1}^* - J_{t}^* \leq - r(\boldsymbol{u}_{0}^*) - \psi_{0}^*$, the cost decreases by at least $1$ at each time step till the task is completed. Considering that $J_t^*$ is always positive and finite, it will take a finite number of steps to reach $J_k^* = 0, \, k\geq t$. Hence, the policy to take the system from $\boldsymbol{w}_{0}$ to $\boldsymbol{w}_{1}$ will be implemented in finite time.

By induction, the system will reach $\boldsymbol{w}_N$ from $\boldsymbol{w}_0$ in finite time.
\end{proof}






 
\section{Numerical Results}

This section shows the efficacy of the proposed EVaR-based risk-sensitive planning method via two numerical examples.

\subsection{Monte-Carlo Simulations}
To illustrate the effectiveness of the proposed method, we compare it to another risk measure, CVaR for different confidence levels, $\alpha$.
We look at the two-dimensional discrete system $ x_{k+1} = Ax_k + Bu_k$, with
\begin{equation*}{\small
   A = \begin{bmatrix} 1.0475 & -0.0463 \\ 0.0463 & 0.9690\end{bmatrix}, \, B = \begin{bmatrix} 0.028 \\ -0.0195\end{bmatrix}.}
\end{equation*}
The control constraints are $$-100 \leq u_k \leq 100.$$ One randomly moving obstacle interferes with the original MPC solution path that would be found in the absence of obstacles. We ran $100$ Monte-Carlo simulations for the two risk measures for different values of the confidence level $\alpha$.

\begin{figure}[tbp]
\vskip -0.1 true in
\centering
        \includegraphics[width=80mm,scale=0.4]{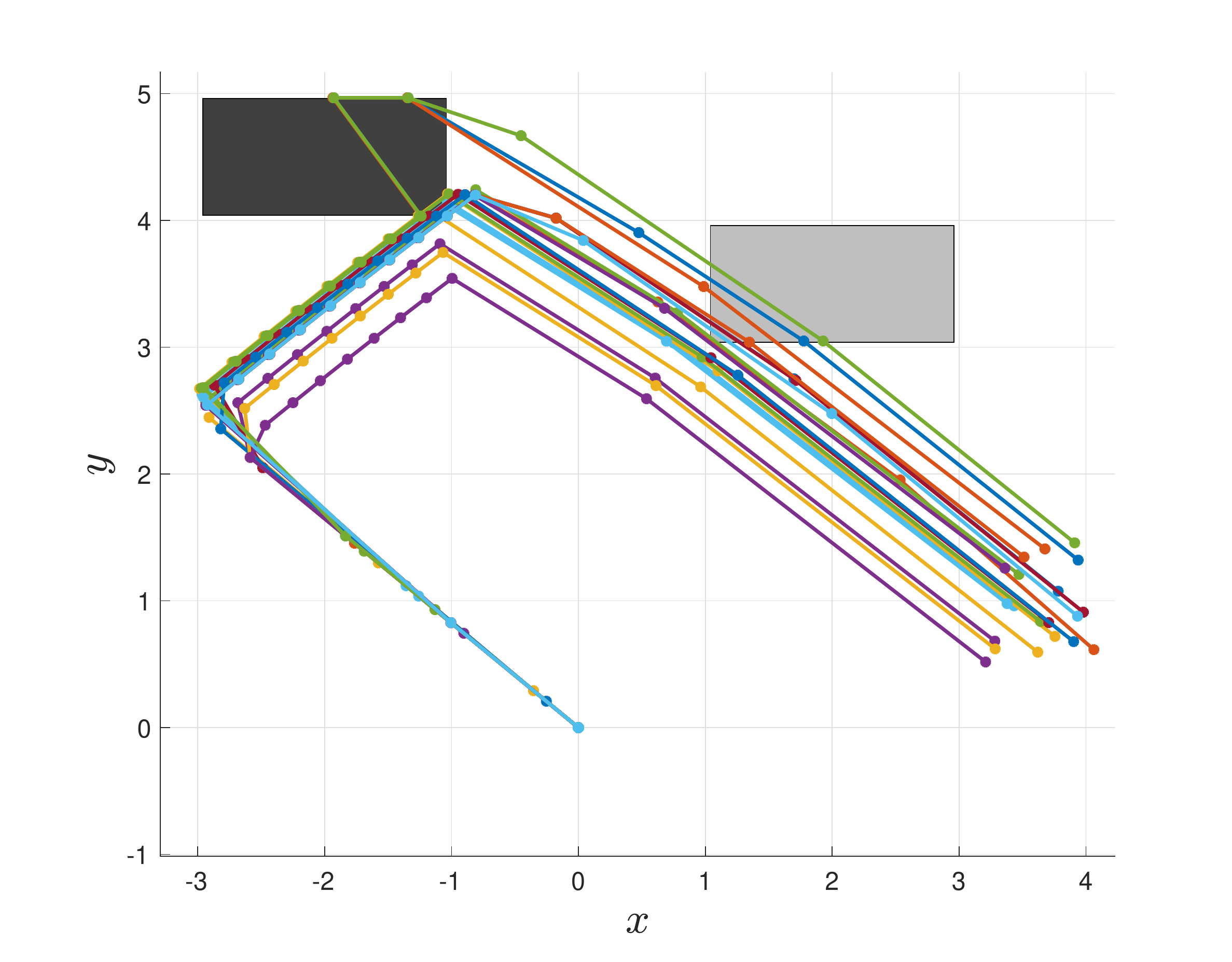}\\
        \vskip -0.2 true in
        \includegraphics[width=80mm,scale=0.4]{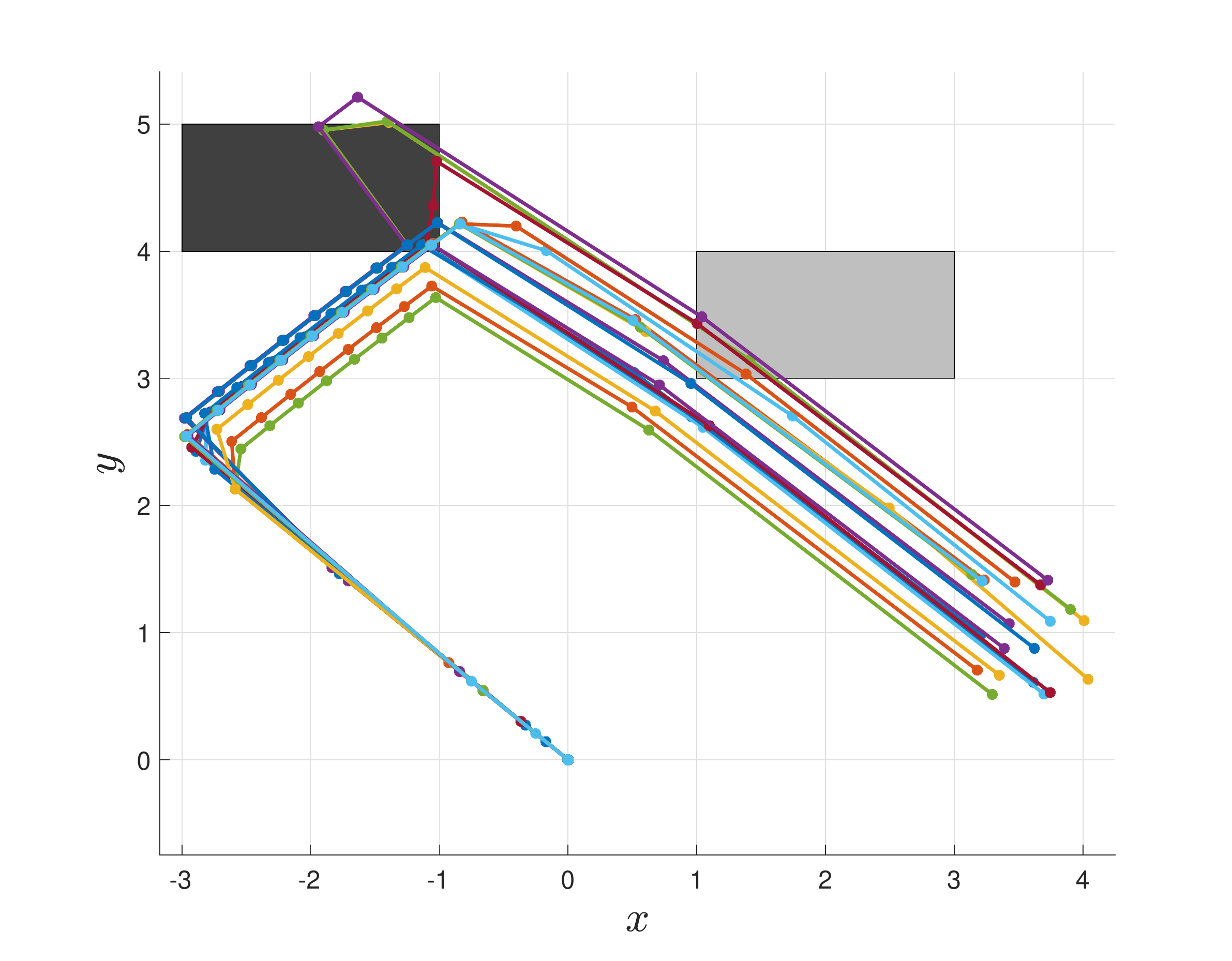}
        \vskip -0.15 true in
\caption{Comparison between EVaR (top) and CVaR (bottom) trajectories. The two uncertain obstacle locations are shown by gray rectangles.  Note that a path may cut through an obstacle between discrete time-steps (which are denoted by a '*').  A collision is not
defined when the system trajectory lies  outside the obstacle at the discrete time-steps. 
}
\label{fig:comp}
\end{figure}

The initial system state, $\boldsymbol{x}_0$, lies somewhere between $(3.1, 0.5)^T$ and $(4.1, 1.5)^T$. For each Monte-Carlo simulation, we randomly chose an initial condition in this range. The paths resulting from this set of initial conditions are most affected by the randomly moving obstacle present at $(-1, 4.5)^T$ with probability $0.75$ and at $(2.5, 3.5)^T$ with probability $0.25$. The risk tolerance is set to $\epsilon = 0.04$. The results are summarized in Table \ref{table:1}. Note that the percentage of collisions is not exact as we run 100 random simulations. A few such trajectories are shown in Fig. \ref{fig:comp} when $\alpha = 0.5$. The two rectangles show both possible obstacle configurations. The darker rectangle has a higher probability of occurrence, $0.75$ and the lighter rectangle has a lower probability of occurrence, $0.25$. The $20$ trajectories seen in the plots are randomly initialized as discussed above. We can see that  more CVaR trajectories intersect the obstacle.
%
\begin{table}[t!]
 \centering
 \begin{tabular}{||c | c c c c c||} 

 \hline
  $\alpha$ &  $0.9$ &  $0.7$ & $0.5$ & $0.3$ & $0.1$\\ [0.5ex] 
 \hline\hline
 EVaR Collisions & $\%0$  & $\%0$ & $\%6$ & $\%3$& $\%66$   \\ 
 \hline
 CVaR Collisions & $\%7$ &  $\%17$ & $\%17$ & $\%14$ & $\%74$\\ 
 \hline
\end{tabular}
\caption{Summary of results from Monte-Carlo simulations}
\label{table:1}
\end{table}

\subsection{Quadcopter}
In this subsection, we implement the waypoint following algorithm - Algorithm \ref{waypoint_alg}. To this end, we consider a quadcopter that must follow given waypoints while avoiding randomly moving obstacles and adhering to state and control constraints. The quadcopter is described by the states $(x, y, z, \phi, \theta, \varphi, \dot{x}, \dot{y}, \dot{z}, \dot{\phi}, \dot{\theta}, \dot{\varphi})^T$. The position of the quadcopter in 3D space is given by the coordinates $x, y, z$ and the roll, pitch, and yaw are given by $\phi, \theta, \varphi$ respectively. The model of the quadcopter is given by 
    \[ \ddot{x} = -g\theta, \, 
       \ddot{y} = g\theta,\, \ddot{z}=-\frac{u_1}{m} - g, \, \]
    \[ \ddot{\phi} = \frac{u_2}{I_{xx}}, \, 
       \ddot{\theta} = \frac{u_3}{I_{yy}}, \, 
       \ddot{\varphi} = \frac{u_4}{I_{zz}},\]
where $m$ is the quadcopter's mass, $g$ is the acceleration due to gravity, and $I_{xx}, I_{yy}, I_{zz}$ are the quadcopter moments of inertia about the $x,y,z$-axes of the system. The control inputs are given by $u_1, u_2, u_3, u_4$. We used the following parameters: $m = 0.65$kg, $I_{xx} = 0.0075$kg.m$^2$, $I_{yy} = 0.0075$kg.m$^2$, $I_{zz} = 0.0013$kg.m$^2$, $g = 9.81$m.s$^{-2}$~\cite{hakobyan2019cvar}.

\begin{figure}[tbp]
\vskip -0.1 true in
\centering
\includegraphics[width=45mm,scale=2]{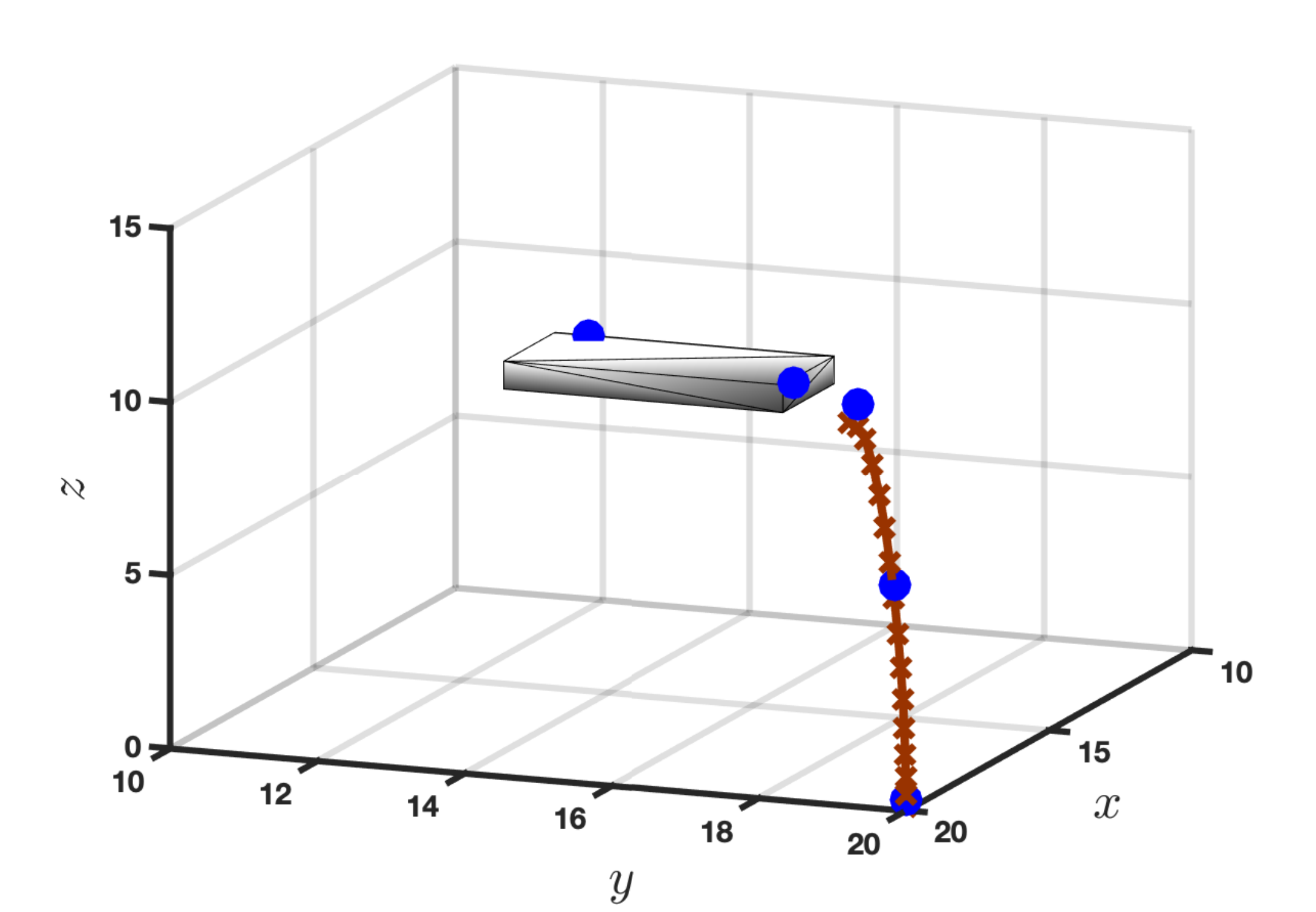}~
        \includegraphics[width=45mm,scale=2]{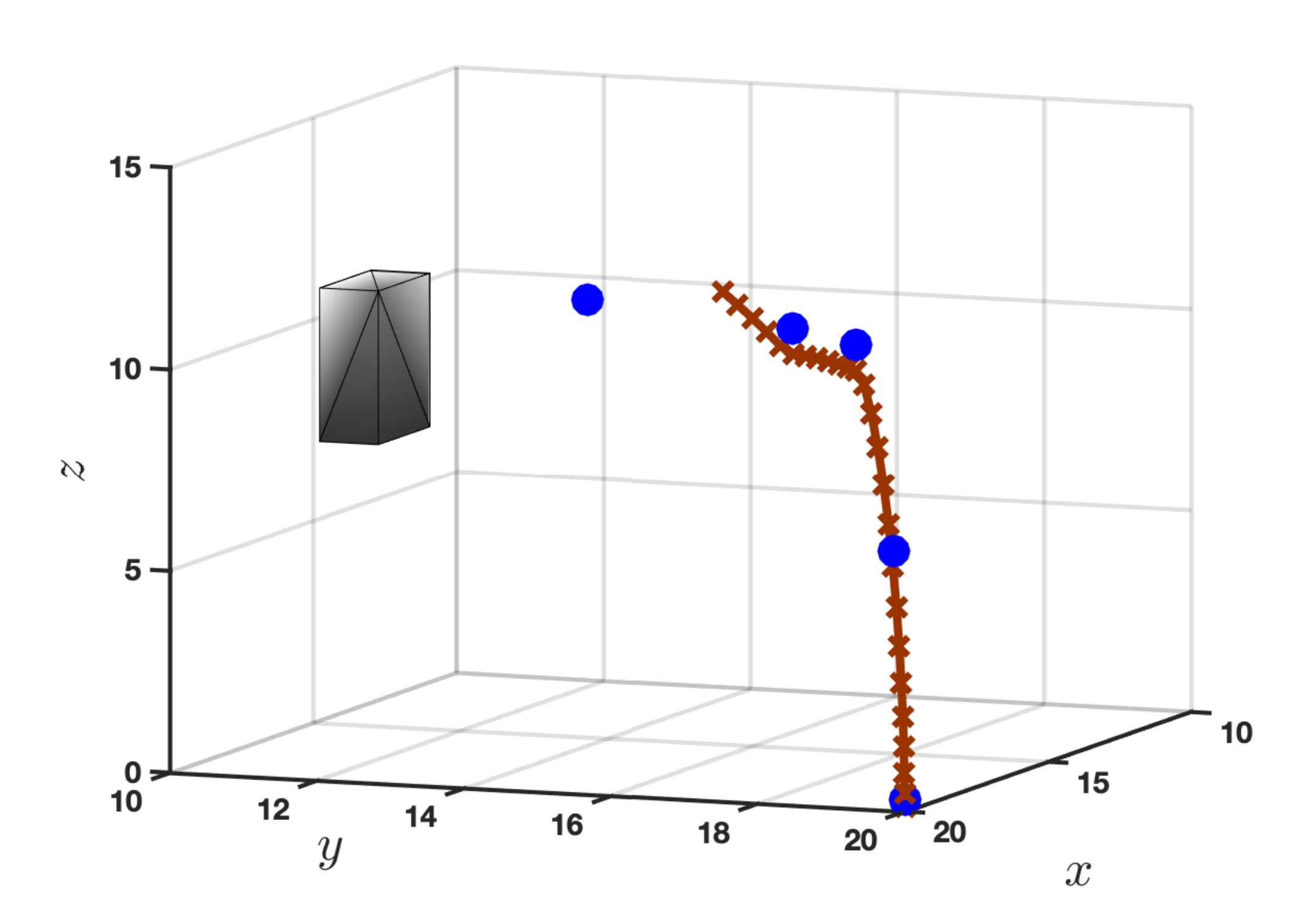}\\
        \includegraphics[width=45mm,scale=2]{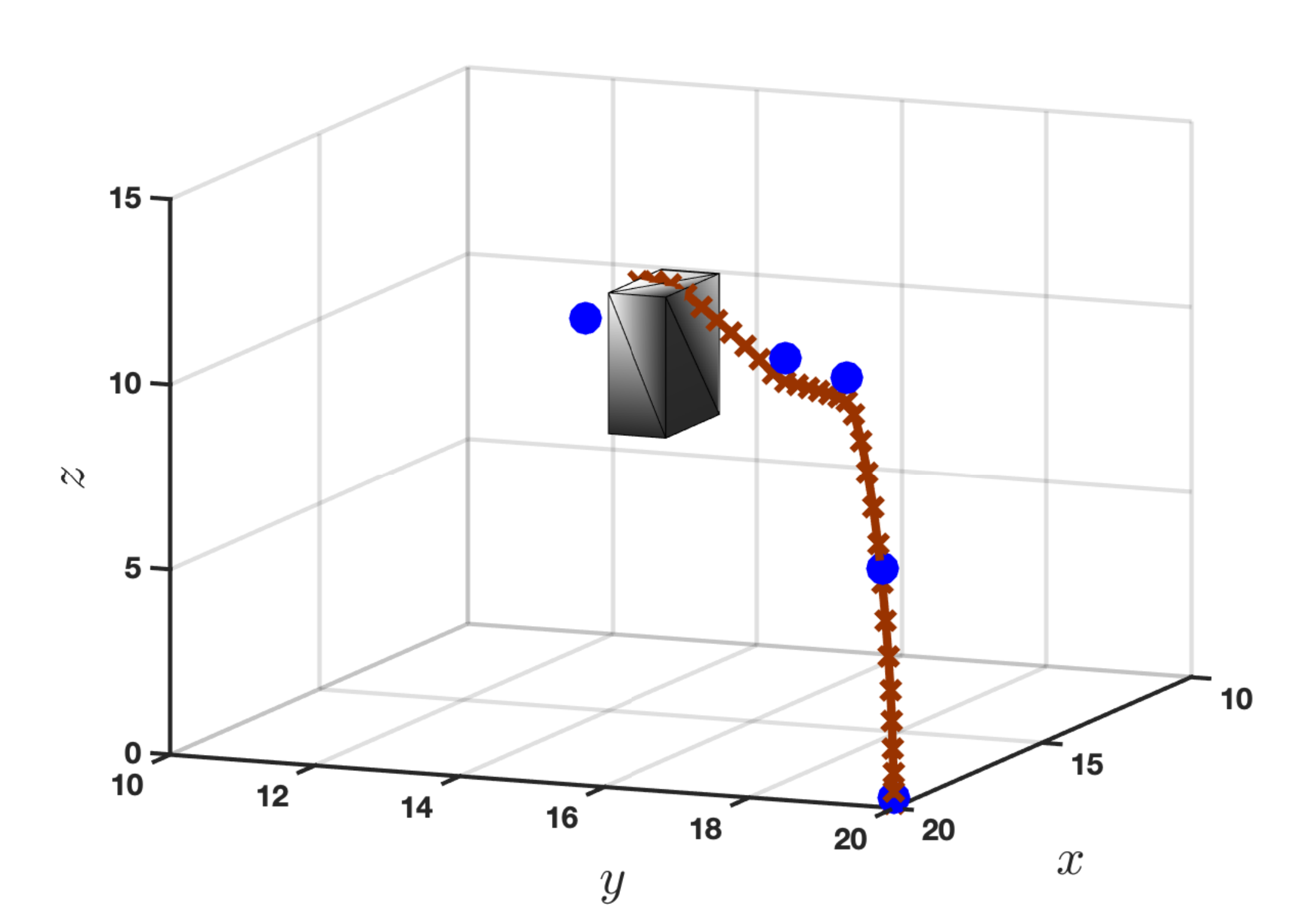}~
        \includegraphics[width=45mm,scale=2]{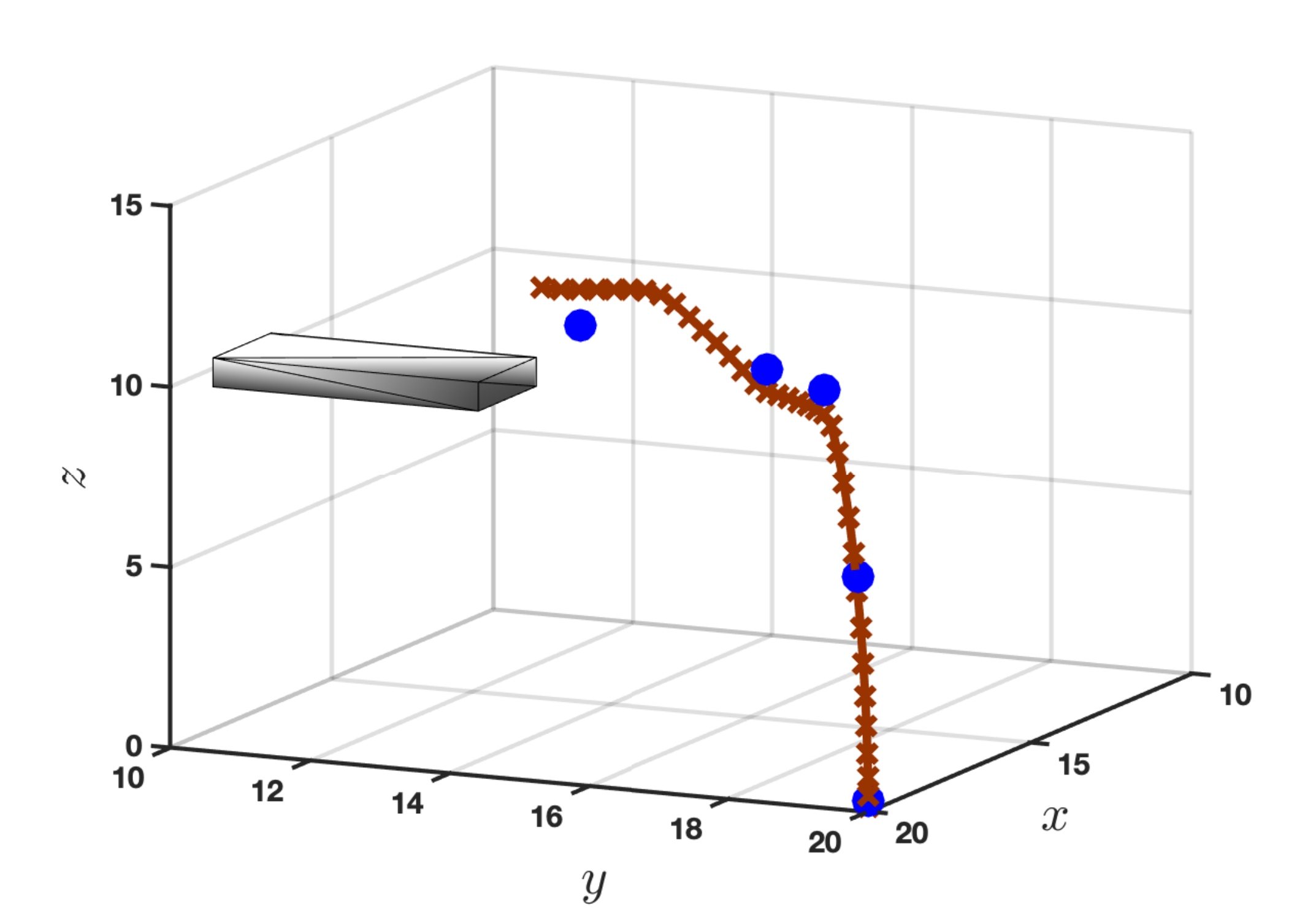}
\caption{Snapshots of the quadcopter trajectory followed using Algorithm \ref{waypoint_alg} (x) given a sequence of waypoints (denoted by o). We can see the obstacle (gray boxes) changes position and orientation with time. The more likely obstacle configuration are shaded darker than the ones less likely to occur. We assume $z = 0$ is the ground plane and gravity pulls the quadrotor in the $-z$ direction.}
\label{fig:waypoints3D}
\end{figure}

The risk constraint has two parameters: the confidence level, $\alpha$, and the risk-threshold, $\epsilon$. We chose $\alpha = 0.5,\, \epsilon = 0.04$. The waypoints are given by regions of size $[-0.5,0.5]^3$ around the waypoint center (denoted by o in Fig. \ref{fig:waypoints3D}). We chose a horizon length of $K = 15$ for the MPC optimization. We considered the case of having one randomly translating and rotating obstacle. The obstacle is a rectangular box of size $2$x$1$x$4$ m$^3$; it can rotate by $\frac{\pi}{2}$ and translate by $4$m along the y-axis and $1$m along the z-axis. Fig. \ref{fig:waypoints3D} shows all the different configurations of this obstacle at different times.



\section{Conclusions}
In this paper, we proposed a risk-constrained motion planning framework for obstacle avoidance. We presented an MPC reformulation in the form of a convex mixed integer program. We showed the recursive feasibility of this optimization and introduced an algorithm to follow waypoints. As shown in this paper, through a comparison with CVaR, the framework is amenable to other risk measures. All coherent risk measures have a convex, bounded, and closed risk envelope. This framework allows any coherent risk measure constrained motion planning problem to be expressed as a convex mixed integer relaxation.

There are many paths of future research for this problem. We could extend this framework to include robust (and by extension risk-sensitive) feasibility to disturbances via constraint tightening \cite{richards2003robustFeasibility}. It is also possible to extend this framework to continuous probability distributions using the relaxation technique that involves sample average approximation \cite{hakobyan2019cvar}. Future work also considers risk-sensitive robot planning under imperfect information~\cite{ahmadi2020pomdp}. 


\balance

\bibliographystyle{plain}
\bibliography{IEEEexample}

\end{document}